\documentstyle[preprint,aps,epsfig,floats]{revtex}
\tightenlines
\draft
\begin{document}
\date{\today}
\preprint{\vbox{\hbox{UA/NPPS-2-02}\hbox{RUB-TPII-03/02}}}
\title{World-line techniques for resumming gluon radiative\\
       corrections at the cross-section level\\}
\author{A.\ I.\ Karanikas${}^{1}$\thanks{E-mail:
        akaranik@cc.uoa.gr},
        C.\ N.\ Ktorides${}^{1}$\thanks{E-mail:
        cktorid@cc.uoa.gr},
        N.\ G.\ Stefanis${}^{2}$\thanks{E-mail:
        stefanis@tp2.ruhr-uni-bochum.de}
        }
\address{${}^{1}$ University of Athens,
                  Department of Physics,
                  Nuclear and Particle Physics Section,   \\
                  Panepistimiopolis,
                  GR-15771 Athens, Greece                 \\
                  [0.1cm]
         ${}^{2}$ Institut f\"ur Theoretische Physik II,
                  Ruhr-Universit\"at Bochum,
                  D-44780 Bochum, Germany                 \\
         }
\maketitle
\begin{abstract}
We employ the Polyakov world-line path-integral version of QCD to
identify and resum at leading perturbative order enhanced
radiative gluon contributions to the Drell-Yan type ($q\bar{q}$
pair annihilation) cross-sections.
We emphasize that this is the first time that world-line techniques
are applied to cross-section calculations.
\end{abstract}
\newpage

\section{Introduction}
\label{sec:intro}
The path, rather than the functional, integral casting of a
relativistic quantum system has a long history, going back to Fock
\cite{Foc37}, Feynman \cite{Fey50}, and Schwinger \cite{Schw51}.
In the course of time it has received substantial contributions from
several authors (see, e.g., Refs.\
\cite{Brink76,Baletal77,KS89,KK92,RSS96,ST02} to name just a few).
However, it was not realized until recent developments in string
theory in the context of effective actions (see
\cite{Bernetal89,Str92,DiVecchiaetal95}, and \cite{Schu01} for a recent
review) that first-quantization methodologies in High Energy Theory can
compete with second-quantization ones.
Of particular interest to us here is the Polyakov world-line path
integral \cite{Pol90}, which employs world-line paths weighted by a
spin factor with the aim to describe the propagation of particle-like
entities in Euclidean space-time.
In fact, Polyakov's intention was to use this construction as a simple
prototype for discussing string quantization.
Hence, for his purposes it was sufficient to consider the simple case
of a free, spin-1/2 particle-like entity.
Motivated by this, two of the present authors \cite{KK95} explored the
possibility of transcribing the matter, spin-1/2, field sector of a
gauge theory into a Polyakov world-line path-integral form.
In these works, it was established, for both Abelian and non-Abelian
gauge systems, that this is, indeed, possible with the spin-factor
making explicitly its entrance in the resulting expression, while the
dynamics enters through a Wilson line (loop) factor defined on each
given path.

Due to the Gaussian character (with respect to the Dirac fields) of the
fermionic sector of physically relevant gauge-field theories, the
aforementioned transcription into a Polyakov world-line path-integral
refers to the full system.
This means that one's way of thinking should be readjusted to the idea
that the second-quantization formalism, associated with the field
theoretical mode of description, can be replaced by a new, but
equivalent, structure that is based on space-time path integrals.
The quantities of central importance defined on these paths are then
the spin-factor and the Wilson line (loop), the latter becoming an
indigenous element of the theory, as it enters at the level of its
definition.

In a number of papers – see, for instance, Refs.\
\cite{KS89,KK95,KKS92,GKKS97} - we have employed the path-integral
casting of either QED or QCD, to study infra-red (IR) factorization
and ensuing behavior of Green's functions and amplitudes in a resummed
perturbative context at the two-, three- and four-point function level.
Roughly speaking, the aforementioned isolation of the long-distance
physics in these theories emerges through the ability to identify a
special set of space-time paths having a very simple geometrical
profile which is shared, in a restricted (but directly relevant to
the physics of the process) neighborhood, by each and every contour
entering the path integral.
In a Euclidean space-time context, the single (multiplicative)
renormalization constant, carried by this special family of paths,
automatically factorizes out \cite{DV80} their contribution to
amplitudes/cross-sections, given that it also accompanies the rest of
the paths.
The more complex geometrical structure of the latter, simply implicates
additional ultraviolet (UV) singularities which can be absorbed into
conventional wave function and coupling constant renormalizations.
This clean, geometrically based, argument, which singularly underlines
the world-line description, will be further elucidated through the main
exposition in the sections to follow.
Minkowski space subtleties, associated with the light cone, which are
encountered in the particular processes under study, will require
separate attention.

Perhaps the most important accomplishment of this paper is that it
extends world-line techniques to cross-section calculations for the
first time.
To be sure, the situation presently considered refers to the
hypothetical situation where the scattering process involves quarks
and associated gluon radiation without reference to hadrons.
It does, nevertheless, fall within the spirit that marks our approach
to IR issues in QCD \cite{KKS92,GKKS97}: Once off-mass shell IR
protection is employed - by an amount that exceeds
$\Lambda_{\rm QCD}$ -
one actually tests how far one can go by remaining strictly within the
confines of QCD before attempting to make contact with real hadrons.
Granted the opposite route, from hadrons to quarks and gluons, via the
use of quantities like structure/fragmentation functions, the
employment of tools such as the operator product expansion, etc.,
constitutes a more realistic procedure for investigating the same
physical problems.
On the other hand, an effort which bases its considerations on a
fundamental theoretical framework in order to arrive at
``cross-sections'' does present merits and interests of its own,
an example of which will be presented below.
In this context, the philosophy underlying our approach to the IR
domain of QCD is closer in spirit to the one articulated by Ciafaloni
in \cite{Cia89}, the only difference being that we shall keep a more
pragmatic (and less ambitious) course by focusing our attention
on cross-section expressions.

Letting these comments suffice for an introductory exposition, we now
proceed to display the organization of the paper, which is as follows.
In the next section, we exhibit the world-line expression for the full
fermionic Green's function and subsequently employ it to construct
corresponding expressions for DY-type QCD amplitudes/cross-sections.
Section \ref{sec:first-order} furnishes, with the aid of an Appendix,
our basic calculations associated with one virtual gluon exchanges for
the specified special set of trajectories.
The resulting expression explicitly reveals the threshold enhancement
factor, whereas the task of virtual gluon resummation is performed,
via the aid of the renormalization group, in Section
\ref{sec:summation}.
Section \ref{sec:realgluon} deals with the resummation of contributions
from real gluon emission.
Finally, in the last section, we further discuss our results and present
our conclusions.

\section{Basic worldline expressions for amplitudes and cross-sections}
\label{sec:basic}
Consider the full two-point (fermionic) Green's function in the
presence of an external gluonic field.
The Polyakov path-integral expression, in Euclidean space-time,
\begin{eqnarray}
  iG_{ij}(x,y|A)
&=&
  \int_{0}^{\infty} dT {\rm e}^{-Tm^{2}}
  \int_{\stackrel{x(0)=x}{x(T)=y}} {\cal D}x(t)
  \left[ m-{1\over 2}\gamma\cdot \dot{x}(T) \right]
  P\exp\left( {i\over 4}\int_{0}^{T} dt
  \sigma_{\mu\nu}\omega_{\mu\nu}
  \right)
\nonumber\\
&& \times
   \exp\left[ -{1\over 4} \int_{0}^{T} dt\dot{x}^{2}(t) \right ]
  P\exp\left[ ig\int_{0}^{T} dt\dot{x}\cdot A(x(t)) \right]_{ij},
\label{eq:green}
\end{eqnarray}
displays the basic world-line features pertaining to this quantity.
Here, and below, $P$ denotes the usual path ordering of the integrals.
The first thing to point out is that a given path of the matter field
quantum, starting at $x$ and ending at $y$ between respective
``proper-time'' values $0$ and $T$, also enters a Wilson line factor.
The latter, being the sole carrier of the dynamics, separates itself
from the rest of the factors in the path integral which are associated
with geometrical properties of paths traversed by spin-1/2 particle
entities.
The most notable such quantity is the so-called spin factor
\cite{Pol90},
$
 P\exp \left[ (i/4)\int_{0}^{T} dt\,\sigma\cdot\omega
       \right ]
$,
where
$
 \omega_{\mu\nu}
=
 (T/2)(\ddot{x}_\mu \dot{x}_\nu-\dot{x}_\mu \ddot{x}_\nu)
$,
accounting, in a geometrical way, for the spin-1/2 nature of the
propagating particle.
Accordingly, our perturbative expansions should be perceived of
in terms of (Euclidean) space-time paths involving a ``proper time''
parameter and {\it not} in terms of Feynman diagrams.
As it turns out \cite{KK99}, in the perturbative context, the
structure of matter particle contours, entering the path integral,
is determined by the points, where a momentum change takes place,
i.e., points where a gauge-field line (real or virtual) attaches
itself on the (fermionic) matter-field path.
The almost everywhere non-differentiability of these contours,
is residing precisely at these points.
A major effort, in this paper, will be devoted to the extension of
the world-line formalism to expressions for cross-sections
corresponding to the particular processes of $q\bar{q}$ annihilation.

From the world-line point of view, the process we intend to study
involves fermionic matter particle (quark) paths that commence at $x$
and end at $y$, being forced to pass through an intermediate point $z$,
where a momentum transfer $Q$ takes place.
This means that the Green's (vertex-type) function we shall be dealing
with has the following form ($\Gamma_\mu$ denotes some Clifford-Dirac
algebra element)
\begin{eqnarray}
  V_{\mu,ij}(y,z,x|A)
&=&
  G_{ik}(y,z|A) \Gamma_{\mu} G_{kj}(z,x|A)
\nonumber\\
&\!=\!&
  \int_{0}^{\infty} dT {\rm e}^{-Tm^{2}}\!\int_{0}^{T} ds \!
  \int_{\stackrel{x(0)=x}{x(T)=y}} {\cal D}x(t)
  \delta \left( x(s)-z \right)
  {\cal G}_{\mu}\left( \dot{x},s \right)
  \exp\left[ -{1\over 4}\int_{0}^{T} dt\,\dot{x}^{2}(t) \right]
\nonumber\\
&&\times
  P\exp\left[ ig\int_{0}^{T}dt\, \dot{x}(t)\cdot A(x(t)) \right]_{ij},
\end{eqnarray}
\label{eq:vertex}
where
\begin{eqnarray}
  {\cal G}_{\mu}\left( \dot{x},s \right)
& \equiv &
\left[ m-{1\over 2}\gamma\cdot \dot{x}(T) \right]
  P\exp\left( {i\over 4}\int_{s}^{T} dt\, \sigma\cdot\omega \right)
  \Gamma_{\mu} \left [ m-{1\over 2}\gamma\cdot \dot{x}(s) \right]
\nonumber\\
&& \times
  P\exp\left( {i\over 4}\int_{0}^{s} dt\, \sigma\cdot\omega\right )
.
\end{eqnarray}
\label{eq:gamma}
It is especially important to realize that in our approach off-shellness
is naturally parameterized in terms of the finite size of the matter
particle contours and realistically accounts for the fact that quarks
reside inside a hadron ($m$ can be viewed as an effective quark mass).

Going over to momentum space, we write
\begin{eqnarray}
  \tilde{V}_{\mu,ij}(p,p'|z|A)
= &&
  \int_{0}^{\infty} dT {\rm e}^{-Tm^{2}} \int_{0}^{T} ds
  \int_{}^{}{\cal D}x(t)\delta \left( x(s)-z \right)
  {\cal G}_{\mu} \left( \dot{x},s \right)
\nonumber\\
&& \!\!\!\!\!\times \;
   \exp\left[ -{1\over 4}\int_{0}^{T} dt\, \dot{x}^{2}(t)
  +ip\cdot x(0)+ip^{\prime}\cdot x(T) \right]
\nonumber\\
&& \!\!\!\!\!\times \;
  P\exp\left[ ig\int_{0}^{T} dt\, \dot{x}(t)\cdot A(x(t)) \right]_{ij}
\nonumber\\
\equiv &&
  \sum_{C^{z}}\tilde\Gamma_{\mu}[C^{z}]P
  \exp \left[ ig\int_{C^{z}}^{} dx\cdot A(x) \right]_{ij},
\label{eq:vertex_mom}
\end{eqnarray}
where $C^{z}$ denotes a generic path forced to pass through point $z$,
at which the momentum $Q$ is imparted.

For a process of the type
$q+\bar{q}\rightarrow$
lepton pair + X the ``amplitude'' expression reads
\begin{eqnarray}
  \Delta_{\mu,ij}
& = &
  \bar{v}(p^{\prime},s^{\prime})(-i\gamma\cdot p^{\prime}+m)
  \tilde{V}_{\mu,ij}(i\gamma\cdot p+m)u(p,s)
\nonumber\\
& \equiv &
  \sum_{C^{z}}\tilde{I}_{\mu,p^{\prime}p}[C^{z}]
  P\exp\left[ ig\int_{C^{z}} dx\cdot A(x) \right]_{ij}
\label{eq:DY_amplitude}
\end{eqnarray}
with the second, comprehensive, expression to be understood
having recourse to Eq.~(\ref{eq:vertex_mom}).

\begin{figure}
\centering\epsfig{file=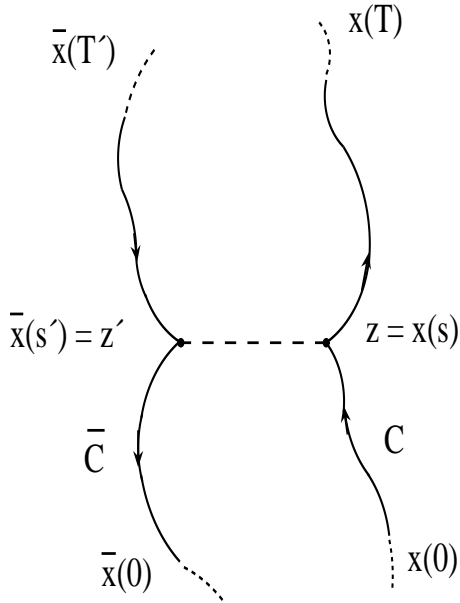,height=8cm,width=6.0cm,silent=}
\vspace{0.5cm}
\caption{\footnotesize
Illustration of two conjugate contours $C$ and $\bar{C}$
entering the world-line path integral, ``talking'' to each
other at points $z$ and $z^{\prime}$, where the momentum
transfer for the physical process takes place.
The distance $|z-z^{\prime}|$ is referred to as the impact
parameter.}
\label{fig:paths1}
\end{figure}

For the cross-section, we need to employ the following quantity,
which we implicitly display in Minkowski space-time after
straightforward adjustments,
\begin{eqnarray}
  \Delta^{\dagger}_\mu \Delta_\nu
& = &
  \sum_{\bar{C}^{z^{\prime}}}
\everymath{\displaystyle}
  \sum_{\rule{0in}{1.4ex}{C^{z}}}
  \tilde{I}^{\dagger}_{\mu,p^{\prime}p}[\bar{C}^{z^{\prime}}]
  \tilde{I}_{\nu,p^{\prime}p}[C^{z}]
\nonumber\\
&& \times
  Tr \left\{
  \bar{P}
  \exp \left[ ig\int_ { \bar{C}^{z^{\prime}} }
  \bar{dx}^{\rho} A_{\rho}(\bar{x}) \right]
  P
  \exp \left[ -ig\int_{C^{z}} dx^{\sigma} A_{\sigma}(x) \right]
     \right\},
\label{eq:crossection}
\end{eqnarray}
where $\bar{P}$ denotes anti-path ordering.
Even though not explicitly displayed, the cross-section acquires a
path-integral form, which has the following characteristics:
\begin{enumerate}
\item Paths $C^{z}$ and $\bar{C}^{z^{\prime}}$ are forced to pass
through points $z$ and $z^{\prime}$, respectively, where the momentum
transfer occurs (see Fig.~\ref{fig:paths1}).
The distance $b\equiv |z-z^{\prime}|$ serves as a measure of how far
apart the two conjugate contours can venture away from each other and
will be referred to as the impact parameter.
\item The traversal of $\bar{C}^{z^{\prime}}$ is made in the opposite
sense relative to $C^{z}$.
If now, we let the two paths join at one end by using translational
invariance, while we allow the other two ends of the contour to close
at infinity, then we obtain the formation of a Wilson loop.
\item Under these circumstances, the Wilson loop formation
guarantees the gauge invariance of the expression for the cross
section.
\end{enumerate}

On the other hand, by keeping the contour lengths finite, but very
large, thereby placing the quarks off-mass-shell, gauge invariance
will still continue to hold to the order of approximation we employ
in our computations, given that the off-mass-shellness serves
at the same time as an IR cutoff.

%

\begin{figure}
\centering \epsfig{file=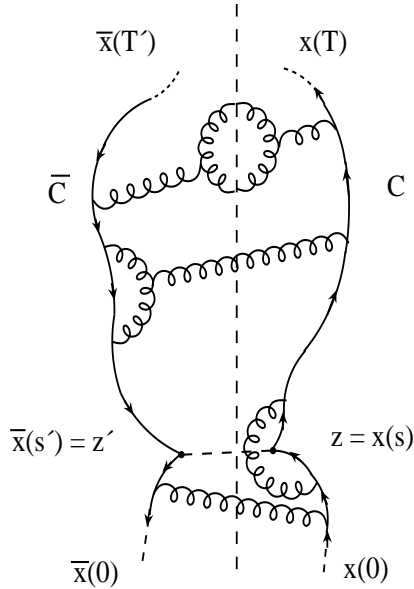,height=8cm,width=6.0cm,silent=}
\vspace{0.5cm}
\caption{\footnotesize
                Virtual gluon radiative corrections of various
                sorts and ``real'' gluon lines with their ends
                attached on each of the two depicted contours at
                the cross-section level.}
\label{fig:paths2}
\end{figure}

Up to this point our considerations have been centered around the
geometrical profile of the paths entering the world-line casting of
QCD, the main conclusion being that, for the process considered, the
relevant contours entering the path integral are marked by a
characteristic point, where a momentum transfer is imparted and that
they are {\it open} for the amplitude and {\it closed} (or almost so)
for the cross-section.
Armed with this information, we now turn our attention to the Wilson
factor which contains all the dynamics of the given process.
The obvious task in front of us is to assess its implications once the
gauge fields are quantized, i.e., once the Wilson factor is inserted
into a {\it functional} integral weighted by the exponential of the
Yang-Mills action.
We display the quantity of interest as follows
\begin{eqnarray}
  {\cal W}
& = &
  \left\langle
        Tr \left\{ \bar{P} \exp\left[ ig\int_{\bar{C}^{z^{\prime}}}
            \bar{dx}^{\mu} A_{\mu}(\bar{x}) \right]
             \right\}_{\rm A}
  \left\{P \exp \left[ -ig\int_{C^{z}}dx^{\nu} A_{\nu}(x)
                \right]
  \right\}_{\rm A}
  \right\rangle
\nonumber  \\
& \equiv &
  \langle Tr(U^{\dagger}(\bar{C}^{z^{\prime}})U(C^{z}))\rangle.
\label{eq:wilson}
\end{eqnarray}
In the above expression, $\{\cdot\cdot\cdot\}_{\rm A}$ signifies
the expectation value with respect to the gauge field functional
integral which, in this work, will be considered in the context of
perturbation theory.
Note in the same context that a virtual gluon attaching itself with
both ends to the fermionic worldline, entering the amplitude,
corresponds to a correlator between a pair of gauge fields originating
from the expansion of the Wilson factor.
On the other hand, for an emitted ``real'' gluon from the fermionic
line, the correlator is between an ``external'' and a Wilson-line
gauge field.\footnote{One will, of course, also encounter correlators
that involve gauge fields from the non-linear terms of the Yang-Mills
action.
These, however, do not enter the leading logarithmic
considerations relevant for our considerations.}
The overall situation is depicted in Fig.~\ref{fig:paths2}.
At the cross-section level, now, ``real'' gluons are integrated with
respect to ``propagators'' linking together the two conjugate
contours, while their polarization vectors are summed over
(cut propagators).
This is precisely what $\langle\cdot\cdot\cdot\rangle$ signifies
in the last equation, as it brackets both Wilson line factors.

This marks a crucial difference to conventional approaches (for example,
\cite{KS95,KM93,Ster87}), wherein the Drell-Yan process is discussed in
a context where IR factorization is based on the eikonal approximation
for soft amplitudes.
Wilson loop expectation values, entering this scheme, are evaluated
along contours corresponding to classical trajectories - along with a
segment which lies on the light cone, introduced in order to secure
gauge invariance.
In our case, by contrast, Wilson contours are built in at a
foundational level, being themselves an integral part of the
description of the full QCD.
Accordingly, factorization properties for us are integrally connected
with the renormalization properties of Wilson loops studied in the
more general context of Refs. \cite{DV80}.
In the light of the above remarks, let us proceed to display
the first-order (in perturbation theory) expression for ${\cal W}$,
which receives contributions from virtual gluons, viz., those
attached at both ends of either the world-line contour $C^z$ or
$\bar{C}^{z^{\prime}}$, as well as from ``real'' gluons linking these
contours to each other (cf. Fig.~\ref{fig:paths2}).
This expression reads
\begin{eqnarray}
  {\cal W}^{(2)}
=
  {\rm Tr}\,I
& - &
     g^{2}C_{\rm F}\int_{0}^{T} dt_{1}
                   \int_{0}^{T} dt_{2}\,
                   \theta \left( t_{2}-t_{1} \right)
                   \dot{x}^{\mu} \left( t_{2} \right)\,
                   \dot{x}^{\nu} \left( t_{1} \right)\,
                   D_{\mu\nu} \left( x(t_{2})-x(t_{1}) \right)
\nonumber\\
&-&
     g^{2}C_{\rm F}\int_{0}^{T^{\prime}}dt^{\prime}_{1}
               \int_{0}^{T^{\prime}}dt^{\prime}_{2}\,
               \theta \left( t^{\prime}_{1}-t^{\prime}_{2} \right)
               \dot{\bar{x}}^{\mu} \left( t^{\prime}_{2} \right)\,
               \dot{\bar{x}}^{\nu} \left( t^{\prime}_{1} \right)\,
\bar{D}_{\mu\nu} \left( \bar{x}(t^{\prime}_{2})-\bar{x}(t^{\prime}_{1})
                 \right)
\nonumber\\
&-&
     g^{2}C_{\rm F}\int_{0}^{T}dt
                   \int_{0}^{T^{\prime}}dt^{\prime}\,
                   \dot{x}(t)\cdot\dot{\bar{x}} \left( t^{\prime}
                                                \right)\,
                   D_{\rm cut} \left( x(t)-\bar{x}(t) \right)
                 + {\cal O}\left(g^{4} \right).
\label{eq:1orderwilson}
\end{eqnarray}
It becomes obvious from their structure that the first two non-trivial
terms correspond to virtual gluon contributions  -- one per
conjugate branch --, while the third one is associated with
``real'' gluon emission.
Finally, concerning the gluon propagators entering the above
equation, we shall be employing their Feynman-gauge form
without loss of generality due to gauge invariance.
In particular we have, in $D$-dimensions,
\begin{equation}
  D_{\mu\nu}(x)
=
  -ig_{\mu\nu}\mu^{4-D}\int\frac{d^{D}k}{(2\pi)^{D}}
  \frac{{\rm e}^{-ik\cdot x}}
  {k^{2}+i0_+}
=
  g_{\mu\nu}{1\over 4\pi^{2}}\left( -\pi\mu^2 \right)^{(4-D)/2}
  \frac{\Gamma(D/2-1)}{(x^{2}-i0_{+})^{(D/2)-1}},
\end{equation}
\label{eq:virtualgluonpropagator}
whereas
\begin{equation}
  D_{\rm cut}(x)
=
  \mu^{4-D}\int\frac{d^{D}q}{(2\pi)^{D}}2\pi\delta(q^{2})\theta(q^{0})
  {\rm e}^{-iq\cdot x}
=
  {1\over 4\pi^{2}}\left( -\pi\mu^{2} \right)^{(4-D)/2}
  \frac{\Gamma(D/2-1)}{\left[ (x_{0}^{2}-i0_{+})^{2}-{\bf x}^{2}
                       \right]^{(D/2)-1}}.
\label{eq:realgluonemission}
\end{equation}
From here on and for the sake of notational simplicity, we shall
simply write $D(x)$ instead of $D_{\rm cut}(x)$.

As already established by other methods, the perturbative
expansion (\ref{eq:1orderwilson}) is plagued by large threshold
logarithms leading to the need for factorization and resummation.
This is precisely the task we are about to undertake within our
framework.

%

\section{First-order virtual gluon corrections in the
vicinities of points \lowercase{$\bbox{z}$} and
\lowercase{$\bbox{z^{\prime}}$}}
\label{sec:first-order}
The space-time mode of description of the Polyakov world-line formalism
puts us into the position to promote the following argument: The point
$z$ (or $z^{\prime}$), where the momentum transfer $Q$ is imparted,
marks the presence of a neighborhood around it, no matter how
infinitesimal in size this might be, whose geometrical structure is
shared by {\it all} fermionic paths entering the path-integral.
Specifically, there will be a derailment (cusp formation), whose
opening angle will be fixed unambiguously, since it is determined by
the momentum transfer.
It follows that the contributions to the amplitude and cross-section
from the immediate vicinities of each of the two cusps is a common
feature of all contours and eventually factorizes.
In this section we shall determine the first-order perturbative term
corresponding exactly to this factor.

Consider now the neighborhood of point $z$ on the contour $C^{z}$.
Expanding around this point, we write
\begin{equation}
  x^{\mu}(t)
=
  x^{\mu}(s) + (t-s) \dot{x}^{\mu}(s\pm 0) + \ldots
\label{eq:expoint}
\end{equation}
with
$v^{\mu}=\dot{x}^{\mu}(s-0)$ and $v^{\prime \mu}=\dot{x}^{\mu}(s+0)$
being entrance and exit four-velocities, respectively, with respect to
$z$.

Adjusting our notation by re-parameterizing the contour so that
the zero value is assigned to point $z$, the relevant
quantity to compute, to first perturbative order, becomes
\begin{eqnarray}
  U_{C,S}^{(2)}
=
  1
&-&
  g^{2}C_{\rm F}
  \left[   \int_{-\sigma}^{0} dt_{1}\int_{-\sigma}^{0} dt_{2}
           \theta \left( t_{2}-t_{1} \right)
           v^{\mu} v^{\nu} D_{\mu\nu} \left( vt_{2}-vt_{1} \right)
         + \int_{0}^{\sigma} dt_{1} \int_{0}^{\sigma} dt_{2}
           \theta \left( t_{2}-t_{1} \right)
\right.
\nonumber\\
&+&
\left.            v^{\prime\mu} v^{\prime\nu}
            D_{\mu\nu} \left( v^{\prime}t_{2}-v^{\prime}t_{1} \right)
          + \int_{0}^{-\sigma} dt_{1}\int_{0}^{\sigma}dt_{2}
            \theta \left( t_{2}-t_{1} \right) v^{\prime\mu} v^{\nu}
            D_{\mu\nu}\left( v^{\prime}t_{2}-vt_{1} \right)
  \right].
\label{eq:U2order}
\end{eqnarray}
It is clear that the above expression corresponds to the first term in
Eq.~(\ref{eq:1orderwilson}), which monitors a virtual gluon
exchange occurring on contour $C$.

From the above considerations it follows that the main contribution
to each double integral comes from the common limit
$t_{1},\,t_{2}\rightarrow 0$.
Suppose now, the other limit is to be determined by demanding that
its contribution to the integrals is of vanishing importance.
Then, such a requirement automatically isolates those contours,
whose only significant geometrical characteristic is that the
four-velocities to approach and depart from point $z$ are
fixed, denoted by $v^{\mu}$ and $v^{\prime\mu}$ respectively.
The same, of course, happens for point $z^{\prime}$, but in the
reverse order.
This justifies the subscript $S$ in $U_{C,S}$, which stands for
``smooth''.
Let us also observe that the omitted terms in Eq.~(\ref{eq:U2order})
will contain negative powers of $\sigma$, whose dimension in the
denominator is (mass)$^2$.
Neglecting their presence means that $\sigma$ should be very large
in magnitude and hence it should be related to an IR cutoff, i.e.,
$\sigma\simeq \lambda^{-2}, \lambda > \Lambda_{\rm QCD}$.

In Euclidean space-time, now, {\it every path} will share the
geometrical structure we are focusing on in some neighborhood of the
point $z$ or ($z^{\prime}$), no matter how close to these points one
has to come.
At the same time, the UV singularities, exhibited by this restricted
set of paths, will entail expressions that solely depend on the two
four-velocities and the opening angle.
Paths of more complex geometrical structure, on the other hand, will
certainly exhibit these UV singularities {\it plus} additional
ones.\footnote{Actually, the standard UV singularities of perturbative
field theories associated with $\beta$-functions, coupling-constant
and wave-function renormalization, pertain to almost everywhere
non-differentiable paths.}
It follows that -- in Euclidean space-time at least -- the restricted
set of trajectories, by exclusively carrying the corresponding
(multiplicative) renormalization constant, {\it factorizes} from the
rest of the expression for the amplitude and/or cross-section.
In a Minkowski space-time context, which will be considered next, we
should anticipate the existence of additional contributions to
Eq.~(\ref{eq:U2order}), due to the light-cone structure that cannot
be assigned to each and every contour and, therefore, cannot be
factorized.
Let us, then, go over to Minkowski space-time, where we have two
distinct possibilities for defining an infinitesimally small
neighborhood around $z$.
The first one, to be labelled $(a)$, reads
\begin{equation}
  (x-x^{\prime})^{2}
=
  {\cal O}(\epsilon^{2}),\,\quad {\rm with} \quad
  v_\mu\simeq v^{\prime}_{\mu},\quad {\rm for}\;\,{\rm all}\quad \mu,
\label{eq:neighbor1}
\end{equation}
where $\epsilon(\leq Q^{-1})$ is a small length scale.
The second alternative, to be labelled (b), can be typically
represented by
\begin{eqnarray}
  (x-x^{\prime})^{2}
&\!\!=\!\!&
  {\cal O}(\epsilon^{2})\,{\rm with}\,
  |v-v^{\prime}|^{2}
=
  {\cal O}(\lambda^{2})\,
  {\rm but}\,(v_{+} - v^{\prime}_{+})
\simeq
  {\cal O}(Q)\,{\rm and}\,(v_{-} - v^{\prime}_{-})
\simeq
  {\cal O}\left( \frac{\lambda^{2}}{Q} \right)
\nonumber\\
&& \quad
  \!\!\!\!\!\!\!\!\!\!\Rightarrow (v_{+}
  - v^{\prime}_{+})(v_{-} - v^{\prime}_{-})
=
  {\cal O}(\lambda^{2})
\label{eq:neighbor2}
\end{eqnarray}
that is equivalently effected via the condition
$v_{+} \gg v^{\prime}_{+},\, v_{-} \simeq v^{\prime}_{-}$.
All in all, there are four different configurations:
$+\leftrightarrow -$ and prime$\leftrightarrow$no-prime
entering this case.

We denote case $(a)$ as ``uniformly soft'', given that
the considered gluon exchanges take place in a neighborhood whose
smallness pertains to all directions.
Case $(b)$, on the other hand, will be referred to as ``jet'' since
gluon emission occurs under circumstances, where entrance and exit
four-velocities differ from each other significantly along one or
the other of the light-cone directions.
Particular implications stemming from this, purely Minkowskian, case as
far as the factorization issue is concerned, will be considered later
on.

Let us commence our calculations by taking up the first
${\cal O}(g^{2})$ term entering the right hand side of
Eq.~(\ref{eq:U2order}).
Since this only involves the branch of the contour $C^{z}$ entering
point $z$, we obtain the same expression regardless of whether or not
a uniformly soft or a jet configuration is being considered.
It reads
\begin{eqnarray}
  I_{1}
& = &
   \int_{-\sigma}^{0} dt_{1} \int_{-\sigma}^{0} dt_{2}\,
   \theta\left( t_{2}-t_{1} \right)v^{\mu} v^{\nu}
  D_{\mu\nu}\left( vt_{2}-vt_{1} \right)
\nonumber\\
&=&
  -\frac{1}{8\pi^{2}}\left( -\pi\mu^{2}L_{1}^{2} \right)^{(4-D)/2}\,
   \Gamma\left( \frac{D}{2}-1 \right)
   \frac{1}{D-3}\,\frac{1}{2-D/2},
\label{eq:I1}
\end{eqnarray}
where\footnote{Note that $v$ has dimensions of mass as our ``time''
parameter $\sigma$ has dimensions of (mass)$^{-2}$.}
$L_{1}= \sigma|v|$.
The second term has the same structure as the first one (it involves
the exiting branch of $C^{z}$) and therefore produces a similar result:
\begin{equation}
  I_{2}
=
  -\frac{1}{8\pi^{2}}\left( -\pi\mu^{2}L_{2}^{2} \right)^{(4-D)/2}\,
   \Gamma\left( \frac{D}{2}-1 \right)
   \frac{1}{D-3}\,\frac{1}{2-D/2}
\label{eq:I2}
\end{equation}
with $L_{2}=\sigma |v^{\prime}|$.

A couple of remarks are in order at this point.
First, even though the length scales $L_1$ and $L_2$ are both large,
being proportional to $\sigma$, they will be of the same order of
magnitude for case (a), whereas for case (b), one scale will be
negligible in comparison with the other.
Accordingly, the total expression to the amplitude for the uniformly
soft contribution will be twice as large as that of the jet-like one.
This being said, we shall denote the dominant length scale by
$L(\simeq L_{1}$ and/or $L_{2})$, when it enters our final expressions,
and set it equal to ${1/\lambda}$, recognizing that it is of the
same order as the IR cutoff.
Second, in order to avoid the double counting resulting from the
fact that each branch has been ``cut-off'' at distance $L$ away from
$z$, where gluon emission occurring at the endpoints will be offset
by a similar one, but opposite in sign, from that portion of the
contour that continues to stretch out to infinity, the final
expressions for the end-point singularities should be multiplied by a
factor of 1/2.
Equivalently, one might think of this compensation as actually
identifying the missing energy of the gluon emission at the
extremities of the path with the off-mass-shellness.
In fact, this is what we have been implying all along when claiming
that finite contours signify off-mass-shellness.

Turning our attention to the contribution resulting from a virtual
gluon exchange from the entrance to the exit branch, with respect to
$z$, we consider the quantity
\begin{eqnarray}
  I_{3}
& = &
  \int_{-\sigma}^{0} dt_{1} \int_{0}^{\sigma} dt_{2}\,
  v^{\prime \mu} v^{\nu}
  D_{\mu\nu}\left( v^{\prime}t_{2}-vt_{1} \right)
=
  \frac{1}{4\pi^2}\left( -\pi\mu^{2} \right)^{(4-D)/2}
  \Gamma\left( {D\over 2}-1 \right)
\nonumber\\
&& \times
  v\cdot v^{\prime}\int_{0}^{\sigma} dt_{1}
  \int_{0}^{\sigma} dt_{2}
  \left(
          t_{1}^{2}v^{2} + t_{2}^{2} v^{\prime 2}
        + 2t_{1}t_{2} v\cdot v^{\prime} - i0_{+} \right)^{1-(D/2)}.
\label{eq:virtglex}
\end{eqnarray}
For case $(a)$ it assumes the form (recall that $v\cdot v^{\prime}$ is
negative)
\begin{eqnarray}
  I_{3}^{(a)}
& = &
  \frac{1}{4\pi^{2}} \left( -\pi\frac{\mu^{2}}{\lambda^{2}} \right)^{(4-D)/2}
  \Gamma\left( {D\over 2}-1 \right)
  \frac{v \cdot v^{\prime}}{|v||v^{\prime}|}
  \int_{0}^{1} dt_{1}
\nonumber \\
&&\times
   \int_{0}^{1} dt_{2}
  \left(  t_{1}^{2}+t_{2}^{2}+2t_{1}t_{2}
  \frac{v \cdot v^{\prime}}{|v||v^{\prime}|}
         -i0_{+} \right)^{1-(D/2)}.
\label{eq:virtglex2}
\end{eqnarray}

As shown in the Appendix, one then determines ($\gamma_{\rm E}$ is
Euler's constant)
\begin{equation}
  I^{(a)}_{3}
=
  \frac{1}{8\pi^{2}}\gamma
  \coth\gamma\frac{1}{2-{D\over 2}}
 +\frac{1}{8\pi^{2}}\gamma
 \coth\gamma
  \ln\left(\frac{\mu^{2}}{\lambda^{2}}
  \pi {\rm e}^{2+\gamma_{\text{E}}}\right),
\label{eq:IDY}
\end{equation}
where
$
 \cosh\gamma
=
 w = -\frac{v\cdot v^{\prime}}{|v||v^{\prime}|}\geq 1
$.

In all of the above expressions, as well as in those that will
follow, we have ignored: (i) all imaginary terms that will drop out
when contributions (for virtual gluons) from the conjugate contour
are taken into account and (ii) finite, $\mu$-independent terms
that will cancel out when real gluon contributions to the
cross-section are included.

Collecting all terms, we deduce, for the ``uniformly smooth'' part,
\begin{equation}
  I_{1}^{(a)}+I_{2}^{(a)}+I_{3}^{(a)}
=
    \frac{1}{8\pi^{2}}(\gamma \coth\gamma-1)
    \frac{1}{2-{D\over 2}}
  + \frac{1}{8\pi^{2}}(\gamma \coth\gamma-1)
    \ln\left(
    \frac{\mu^{2}}{\lambda^{2}}\pi {\rm e}^{2+\gamma_{\text{E}}}
    \right).
\label{eq:I123}
\end{equation}

Concerning the ``jet'' part of the computation, we only need
to consider $I_{3}^{(b)}$ because \footnote{Recall the
remark following Eq.~(\ref{eq:I2}).} the expression for
$I_{1}^{(b)}+I_{2}^{(b)}$ is simply one half of that of
$I_{1}^{(a)}+I_{2}^{(a)}$.
A typical term entering
$I_{3}^{(b)}$ $(v_{-} \gg v^{\prime}_{+})$
is
\begin{equation}
  I_{3}^{(b)}
=
  \frac{1}{4\pi^{2}}\left( -\pi\mu^{2} \right)^{(4-D)/2}
  \Gamma\left( {D\over 2}-1 \right) v\cdot v^{\prime}
  \int_{0}^{\sigma} dt_{1} \int_{0}^{\sigma} dt_{2}
  \left( t_{1}^{2}v^{2}+2t_{1}t_{2}v\cdot v^{\prime}-i0_{+}
  \right)^{1-(D/2)},
\label{eq:I3}
\end{equation}
whose computation suffices to furnish each of the other three terms
as well.

It is shown in the latter part of the Appendix that one obtains
\begin{equation}
  I_{3}^{(b)}
=
    \frac{1}{16\pi^{2}}\,\frac{1}{\left( 2-{D\over 2} \right)^{2}}
  + \frac{1}{16\pi^{2}}\,\frac{1}{2-{D\over 2}}
    \ln\left( \frac{\mu^{2}}{\lambda^{2}}\pi {\rm e}^{\gamma_{\rm E}}
       \right)
  + \frac{1}{32\pi^{2}}
    \ln ^{2}\left( \frac{\mu^{2}}{\lambda^{2}}
    \pi {\rm e}^{\gamma_{\text{E}}} \right)
  + \mbox{const}.
\label{eq:I3fin}
\end{equation}
It is duly observed that the singularity structure of the above
expression is $\gamma$-independent.
In fact, the ``jet'' configuration is a Minkowski-space feature and is
connected to ``gluon'' emission in the + or the - light-cone
coordinates direction.
This result is in accord with Wilson loop studies in Minkowski space,
wherein the relevant contour lies partly on the light cone \cite{KM93}.
Subtracting the pole terms in the $\overline{{\rm MS}}$ scheme, we
arrive at the finite part of the overall result.
For the uniformly soft contribution, in particular, we get
\begin{equation}
  (I_{1}^{(a)}+I_{2}^{(a)}+I_{3}^{(a)})_{\text{fin}}
=
  \frac{1}{8\pi^{2}}
\everymath{\displaystyle}
  \left( \gamma \coth\gamma-1 \right)
  \ln\left( \frac{\mu^{2}}{{\rule{0in}{2.0ex}\bar{\lambda}}^{2}}
     \right),
\label{eq:I123finsoft}
\end{equation}
while the jet contribution reads
\begin{equation}
  (I_{1}^{(b)}+I_{2}^{(b)}+4I_{3}^{(b)})_{\text{fin}}
=
\everymath{\displaystyle}
  \frac{1}{8\pi^{2}}
  {\ln}^{2}\left( \frac{\mu^{2}}{{\rule{0in}{2.0ex}\bar{\lambda}}^{2}}
           \right)
  - \frac{1}{16\pi^{2}}\ln \frac{\mu^{2}}{\bar{\lambda}^{2}},
\label{eq:I123finjet}
\end{equation}
where we have set
$\bar{\lambda}^{2}\equiv 4\lambda^{2} {\rm e}^{-2\gamma_{\rm E}}$.
The above relation takes into account all four different
configurations contributing to $I_{3}^{(b)}$.

Gathering all terms, we arrive at the following overall result for
the second-order contribution stemming from contour $C^{z}$
\begin{equation}
\everymath{\displaystyle}
  U^{(2)}_{C,S}
=
  1 - \frac{\alpha_{\text{s}}}{2\pi} C_{\rm F}
  \left[ \left( \gamma \, \coth \gamma -1 \right)
          \ln \left( \frac{\mu^{2}}
                          {{\rule{0in}{2.0ex}\bar{\lambda}}^{2}}
              \right)
      - {1\over 2}
        \ln \left( \frac{\mu^{2}}{{\rule{0in}{2.0ex}\bar{\lambda}}^{2}}
                         \right)
        + \ln ^{2} \left(
                   \frac{\mu^{2}}{{\rule{0in}{2ex}\bar{\lambda}}^{2}}
                   \right)
  \right].
\label{eq:UfinCz}
\end{equation}
A similar result is obtained also for contour $\bar{C}^{z^{\prime}}$.

Noting that
$\gamma \coth \gamma = \ln \left( Q^{2}/m^{2} \right)$
(for $Q^{2}\gg m^{2}$),
with
$ Q^{2} = (p+p^{\prime})^{2} $,
we recognize that the well-known perturbative enhancements occurring
as
$Q^{2} \rightarrow \infty$
are associated with the eikonal-type trajectories upon which our
present calculations have been based.
One, now, realizes that these trajectories define {\it threshold}
conditions, with respect to the given momentum exchange $Q$, for the
process under consideration, since they leave no room for space-time
contour fluctuations.
In the following section, we shall treat the resummation of these
enhanced contributions to leading logarithmic order.
We shall, furthermore, identify a correction factor associated
with those terms in Eq.~(\ref{eq:UfinCz}) not involving the
enhancement factor $\ln(Q^{2}/m^{2})$.

\section{Resummation of enhanced contributions from virtual gluons}
\label{sec:summation}
The family of world-line paths to which the considerations in the
previous section refer was used in order to deal with all (virtual)
single-gluon exchanges, consistent with the simple geometrical
configuration of two constant four-velocities making a fixed angle
$\gamma$ between them (in Euclidean formulation).
Among these gluons there will be ``hard'' ones (upper limit $Q$) and
``soft'' ones (lower limit set by $\bar{\lambda})$.
What is debited to the former and what to the latter group of gluons
is, of course, relative.
It is precisely the role of the renormalization scale $\mu$, entering
through the need to face UV divergences arising even for the
restricted family of paths, to provide the dividing line.
The corresponding renormalization-group equation reflects the fact
that the scale $\mu$ is arbitrary and that physical results do not
depend on it.
A straightforward application of this fact will enable us to resum
the enhanced, virtual gluon contribution to the amplitude in leading
logarithmic order, as well as to obtain a bona-fide correction term.

To bring the above discussion into a concrete form, let us first
consider a separation, good to order $1/Q^{2}$, of the cusp
contribution, which can be factorized from the amplitude $U_{C}$,
entering Eq.~(\ref{eq:wilson}), on the basis of what has been
determined so far.
Hence, we write
\begin {equation}
\everymath{\displaystyle}
  U_{C}
=
  U_{C,{\rm cusp}}\left ({Q^{2}\over m^{2}},{\mu^{2}\over
  {\rule{0in}{2ex}\bar{\lambda}}^{2}}
         \right)
  \hat{U}_{C}\left( {Q^{2}\over \mu^{2}},
  {\mu^{2}\over {\rule{0in}{2ex}\bar{\lambda}}^{2}}
  \right)+{\cal O}\left( {1\over Q^{2}} \right)
\label{eq:Ufact}
\end{equation}
with
\begin{equation}
\everymath{\displaystyle}
  U_{C,\text{cusp}}^{(2)}
=
  1-\frac{\alpha_{\text{s}}}{2\pi}C_{\rm F}
  \left( \gamma\coth\gamma -1  \right)
  \ln\left( {\mu^{2}\over {\rule{0in}{2ex}\bar{\lambda}}^{2}} \right),
\label{eq:Ucusp}
\end{equation}
where we have normalized $U_{C,\text{cusp}}$ to unity for
$\gamma \to 0$.
The designation ``cusp'', above, refers to that factor of the soft
sector, which recognizes the angle $\gamma$.
The factor $\hat{U}_{C}$, on the other hand, includes {\it both}:
(i) soft contributions - related to the dependence on the quantity
${\mu^2/\bar{\lambda}^2}$ - {\it and} (ii) hard ones - depending on
the quantity ${Q^2/\mu^2}$.

It is convenient to take the logarithmic derivative of
Eq.~(\ref{eq:Ufact}) with respect to $Q^2$:
\begin{equation}
  \frac{d}{d\ln Q^{2}}\ln U_{C}
=
    \frac{d}{d\ln Q^{2}}\ln U_{C,\text{cusp}}
  + \frac{d}{d\ln Q^{2}}\ln \hat{U}_{C}
  + {\cal O}\left( {1\over Q^{2}} \right).
\label{eq:noUcoll}
\end{equation}

The $\mu$-independence of physical results leads to the renormalization
group equation whose ultimate justification has to do with the
multiplicative renormalization of the soft (cusp-angle dependent)
factor.
Indeed, the latter is detached from collinear emission and totally
complies with the Euclidean space-time properties of Wilson loops for
which the results of Refs.\ \cite{DV80} fully apply.
Specifically, we write
\begin{equation}
  \frac{d}{d\ln\mu}\frac{d}{d\ln Q^{2}}\ln \hat{U}_{C}
=
  -\frac{d}{d\ln\mu}\frac{d}{d\ln Q^{2}}
   \ln U_{C,\text{cusp}}
=
  \Gamma_{\text{cusp}}(\alpha_{\text{s}})
\label{eq:muindep}
\end{equation}
with $\Gamma_{{\rm cusp}}$ to be read off from
Eqs.~(\ref{eq:I123})-(\ref{eq:I3fin}) and (\ref{eq:U2order}):
\begin{equation}
  \Gamma_{\text{cusp}}(\alpha_{\text{s}})
=
  {\alpha_{\text{s}}\over\pi}C_{\rm F}+{\cal O}(\alpha_{\text{s}}^{2}).
\label{eq:Gammacusp}
\end{equation}

From the second leg of Eq.~(\ref{eq:muindep}), one obtains
\begin{equation}
  \frac{d}{d\ln Q^{2}}\ln U_{C,\text{cusp}}
=
  -\int_{\bar{\lambda}^{2}}^{\mu^{2}}\frac{dt}{2t}
   \Gamma_{\text{cusp}}\left[ \alpha_{\text{s}}(t) \right]
\label{eq:cuspcusp}
\end{equation}
which, in turn, gives
\begin{equation}
  \frac{d}{d\ln Q^{2}}\ln \hat{U}_{C}
=
  -\int_{\mu^2}^{Q^2}\frac{dt}{2t}
  \Gamma_{\text{cusp}}\left[ \alpha_{\text{s}}(t) \right]+
  \Gamma\left[ \alpha_{\text{s}}(Q^{2}) \right],
\label{eq:cuspH}
\end{equation}
where we have defined
\begin{equation}
  \frac{1}{2}\Gamma\left[ \alpha_{\text{s}}(Q^{2}) \right]
\equiv
  \frac{d}{d\ln Q^{2}}\ln \hat{U}_{C}
  \left( {Q^{2}\over\mu^{2}},{\mu^{2}
         \over {\rule{0in}{2ex}\bar{\lambda}}^{2}}
  \right)_{\mu^{2}=Q^{2}}.
\label{eq:defGamma}
\end{equation}

Combining the last three equations, we have
\begin{equation}
  \frac{d}{d\ln Q^{2}}\ln U_{C}
=
  -\int_{\bar{\lambda}^{2}}^{Q^{2}}\frac{dt}{2t}
     \Gamma_{\text{cusp}}\left[ \alpha_{\text{s}}(t) \right]
   + \frac{1}{2}\Gamma\left[ \alpha_{\text{s}}(Q^{2}) \right],
\label{eq:Gammatot}
\end{equation}
where, in terms of $\ln U_{C}$, we write
\begin{equation}
  \frac{1}{2}\Gamma\left[ \alpha_{\text{s}}(Q^{2}) \right]
\equiv
  \frac{d}{d\ln Q^{2}}\ln U_{C}|_{\bar{\lambda}^{2}=Q^{2}}.
\label{eq:Gamma}
\end{equation}

Setting $\mu^{2}=Q^{2}$ in Eq.~(\ref{eq:Ufact}),
we are led to the identification
\begin{equation}
  \frac{1}{2}\Gamma\left[ \alpha_{\text{s}}(Q^{2}) \right]
=
     \frac{d}{d\ln Q^{2}}
     \ln U_{C,\text{cusp}}\left( \frac{Q^{2}}{m^{2}},
     \frac{Q^{2}}{\bar{\lambda}^{2}} \right )
     |_{\bar{\lambda}^{2}=Q^{2}}
   + \frac{d}{d\ln Q^{2}}
   \ln \hat{U}_{C}
   \left( 1, \frac{Q^{2}}{\bar{\lambda}^{2}}
   \right)\Biggr|_{\bar{\lambda}^{2}=Q^{2}}.
\label{eq:GammaUcoll}
\end{equation}
To second order we have
\begin{equation}
  \Gamma^{(2)}\left[ \alpha_{\text{s}}(Q^{2}) \right]
=
  {3\over 2}\frac{C_{\rm F}}{\pi}\alpha_{\text{s}}(Q^{2})+
  {\cal O}(\alpha_{\text{s}}^{2}).
\label{eq:Gamma2}
\end{equation}
Gathering our findings, we obtain our final, resummed
result corresponding to the contour $C$.
It reads
\begin{equation}
  U_{C}
=
  \exp\left\{-\int_{\bar{\lambda}^{2}}^{Q^{2}}\frac{dt}{2t}
       \left[ \ln{Q^{2}\over t}\Gamma_{\text{cusp}}(\alpha_{\text{s}}(t))
      -\Gamma(\alpha_{\text{s}}(t))
      \right]
      \right\}
  U_{C,0}(\alpha_{\text{s}}(Q^{2})).
\label{eq:resU}
\end{equation}
One notes that the second (``correction'') term in the square brackets
is associated with collinear emission (cf. Eq.~(\ref{eq:GammaUcoll})).
Finally, the factor
$U_{C,0}(\alpha_{s}(Q^{2}))$
represents input from initial conditions {\it at the QCD level}.
Clearly, the conjugate-contour term
$U^{\dagger}(\bar{C}^{z^{\prime}})$
can be treated in a completely analogous fashion.

\section{Resummation of enhanced contributions from real gluon
         emission}
\label{sec:realgluon}
\begin{figure}
\centering \epsfig{file=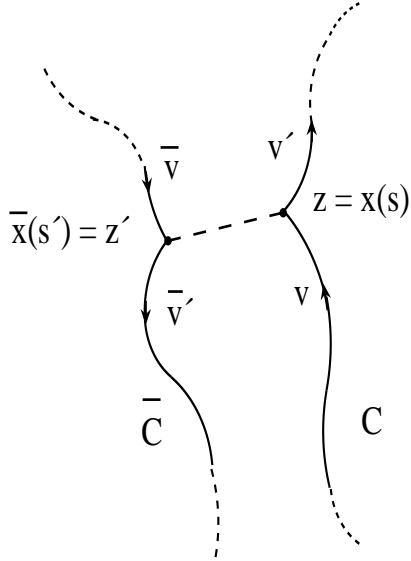,height=8cm,width=6.0cm,silent=}
\vspace{0.2cm}
\caption{\footnotesize
                Neighborhoods of respective points on two
                conjugate contours, where the momentum transfer
                takes place, and associated four-velocities.}
\label{fig:paths3}
\end{figure}

We shall now turn our attention to real gluons and attempt to
factorize cross-section contributions from neighborhoods around
points $z$ and $z^{\prime}$.
Note that this time we have to deal with gluons which connect two
``opposite'' neighborhoods while crossing the unitarity line
(this situation is depicted in Fig.~\ref{fig:paths3}).
The relevant scale promptly entering our considerations is the impact
parameter $b=z-z^{\prime}$, which must be eventually integrated over
in order to get the physically measurable cross-section.
Naturally, the short-distance cutoff in this integration will be
provided by the (length) scale $1/|Q|$.

%

For the eikonal-type family of paths, and in first-order perturbation
theory, the relevant quantity on which our quantitative considerations
are to be based, i.e., the counterpart of Eq.~(\ref{eq:U2order}),
is given by
\begin{eqnarray}
  U^{(2)}_{C\bar{C},{\rm S}}
=
  1
& + &
  g^{2}C_{\rm F}\left[
                      \int_{-\sigma}^{0} dt_{1}
                      \int_{-\sigma}^{0} dt_{2}
                      v\cdot\bar{v}\, D\left( t_{1}v-t_{2}\bar{v}+b
                                       \right)\right.
                      \nonumber \\
& + & \left.
                      \int_{0}^{\sigma} dt_{1}
                      \int_{0}^{\sigma} dt_{2}
                      v^{\prime}\cdot\bar{v}^{\prime}\,
                      D\left( t_{1}v^{\prime}-t_{2}\bar{v}^{\prime}+b
                      \right)\right.
\nonumber\\
& + & \left.
                      \int_{-\sigma}^{0} dt_{1}
                      \int_{0}^{\sigma} dt_{2} v\cdot\bar{v}^{\prime}\,
                      D\left( t_{1}v-t_{2}\bar{v}^{\prime}+b
                      \right)\right.
\nonumber \\
& + & \left.
                      \int_{0}^{\sigma} dt_{1}
                      \int_{-\sigma}^{0} dt_{2} v^{\prime}\cdot\bar{v}\,
                      D\left( t_{1}v^{\prime}-t_{2}\bar{v}+b
                      \right)
               \right],
\label{eq:U2orderreal}
\end{eqnarray}
where the bar denotes four-velocities for the conjugate contour and
the subscript cut is henceforth omitted.

To identify the leading behavior of $U^{(2)}_{C\bar{C},S}$,
with respect to $b$, we shall consider first the situation
corresponding to $b=0$.
The subsequent emergence of UV divergences, once handled through
dimensional regularization, will introduce a mass scale $\mu^{\prime}$
that will be bounded from below by an IR cutoff $\lambda$ and from
above by the (mass) scale $1/b$.
The resulting renormalization group equation will facilitate the
resummation of the leading terms, just as in the virtual-gluon case.

Let us start with our quantitative considerations by looking at the
term
\begin{equation}
  J_{1}(b)
\equiv
  v\cdot\bar{v} \int_{-\sigma}^{0} dt_{1}
  \int_{-\sigma}^{0} dt_{2}\, D\left( t_{1}v-t_{2}\bar{v}+b \right)
\label{eq:Jb}
\end{equation}
with $v^2=\bar{v}^2= -v\cdot\bar{v}$ (see Fig.~\ref{fig:paths3}).

Setting $b=0$ and using the expression for the cut propagator as given
by Eq.~(\ref{eq:realgluonemission}), we obtain
\begin{equation}
  J_{1}(0)
=
  -{1\over 4\pi^{2}}
  \left( -\pi\mu^{\prime 2}L_{1}^{2} \right)^{\left(2-D/2\right)}
  \Gamma\left({D\over 2}-1\right)\,{1\over D-3}\,{1\over 4-D}
  \left[ 1-(2^{4-D}-1) \right]
\label{eq:J0}
\end{equation}
which actually coincides with what one would obtain if the regular
propagator was substituted.
The significance of this occurrence is that it leads to the same
anomalous dimensions for the running of the real-gluon contribution
to the cross-section as for the virtual part.
This fact can be immediately verified via a direct comparison
with Eq.~(\ref{eq:I1}).

Isolating the finite part of the above expression, we write
\begin{equation}
  J_{1}^{(a),\text{fin}}
=
  -{1\over 8\pi^{2}} \ln\left( \mu^{\prime 2}\over\lambda^{2}\right).
\label{eq:J1fin}
\end{equation}
It is trivial to see that the same result holds also for
$J_{2}^{(a),{\rm fin}}$.

We next turn our attention to the term
\begin{equation}
  J_{3}(b)
\equiv
     v\cdot\bar{v}^{\prime} \int_{-\sigma}^{0} dt_{1}
     \int_{0}^{\sigma} dt_{2}\,D\left( t_{1}v-t_{2}\bar{v}^{\prime}
   + b \right).
\label{eq:J3b}
\end{equation}
Its computation will concurrently allow us to determine $J_{4}(b)$,
which corresponds to the exchange prime$\leftrightarrow$ no-prime in
the expression above.

Dimensionally regularizing the cut propagator, we then obtain
\begin{eqnarray}
  J_{3}(0)
& = &
  {1\over 4\pi^{2}}(-\pi \mu^{\prime 2})^{(4-D)/2}
  \Gamma\left( {D\over 2}-1 \right)
  v\cdot\bar{v}^{\prime}\!\!\int_{0}^{\sigma} dt_{1}
\nonumber \\
&& \times
  \int_{0}^{\sigma}\! dt_{2}
  \left( t_{1}^{2}v^{2}+t_{2}^{2}\bar{v}^{\prime
  2}+2v\cdot\bar{v}^{\prime}t_{1}t_{2}
  -i0_{+} \right)^{1-D/2}.
\label{eq:cutprop}
\end{eqnarray}
Once again we record, by referring to Eq.~(\ref{eq:virtglex}),
coincidence of the singularities and, by extension, of associated
anomalous dimensions between virtual and real gluon expressions that
contribute to the cross section.

For the ``uniformly soft'' configuration the corresponding result is
\begin{eqnarray}
  J_{3}^{(a)}(0)
& = &
  {1\over 4\pi^{2}}(-\pi \frac{\mu^{\prime 2}}{\lambda^{2}})^{(4-D)/2}
  \Gamma\left( {D\over 2}-1 \right)\!
  \frac{v\cdot\bar{v}^{\prime}}{|v||\bar{v}^{\prime}|}
\nonumber \\
&& \times
  \int_{0}^{1}\! dt_{1}\!
  \int_{0}^{1}\! dt_{2}\!
  \left( t_{1}^{2}+t_{2}^{2}+2t_{1}t_{2}
  \frac{v\cdot\bar{v}^{\prime}}{|v||\bar{v}^{\prime}|}-i0_{+}
  \right)^{1-D/2}.
\label{eq:Jsoft}
\end{eqnarray}

Taking into consideration that
$
 \frac{v\cdot\bar{v}^{\prime}}{|v||\bar{v}^{\prime}|}
=
 \frac{v^{\prime}\cdot\bar{v}}{|v^{\prime}||\bar{v}|}
=
 \cosh\gamma > 0
$
we obtain
\begin{eqnarray}
  J_{3}^{(a)}(0)
=
  J_{4}^{(a)}(0)
& = &
  {1\over 4\pi^{2}}(-\pi \frac{\mu^{\prime 2}}{\lambda^{2}})^{(4-D)/2}
  \Gamma\left( {D\over 2}-1 \right)
  \cosh\gamma
\nonumber\\
&& \times
  \int_{0}^{1} dt_{1}
  \int_{0}^{1} dt_{2} \,
  \left( t_{1}^{2}+t_{2}^{2}+2t_{1}t_{2}
  \cosh\gamma-i0_{+} \right)^{1-D/2},
\label{eq:J3DY}
\end{eqnarray}
whose finite part reads
\begin{equation}
  J_{3}^{(a),\text{fin}}(0)
=
  J_{4}^{(a),\text{fin}}(0)
=
  {1\over 8\pi^{2}}
  \gamma\coth\gamma\,
  \ln\left( {\mu^{\prime 2}\over\lambda^{2}} \right).
\label{eq:J3DYfin}
\end{equation}

Turning now our attention to the ``jet'' configuration, we can actually
go directly to
$J_{3}^{(b)}(0)$,
since
$J_{1}^{(b)}(0)+J_{2}^{(b)}(0)$
furnishes half of the contribution of its uniformly soft counterpart,
the reason being the same as the one given in the virtual gluon case.
We thus have
\begin{eqnarray}
  J_{3}^{(b)}(0)
= &&
    {1\over 4\pi^{2}}\left( -\pi\mu^{\prime 2} \right)^{(4-D)/2}
  \Gamma\left( {D\over 2}-1 \right)
  \frac{v\cdot\bar{v}^{\prime}}{|v|}
\nonumber \\
&& \times
  \int_{0}^{1} dt_{1}
  \int_{0}^{1} dt_2\, \left(t_{1}^{2}+2t_{1}t_{2}
                            \frac{v\cdot\bar{v}^{\prime}}{|v|}-i0_{+}
                      \right)^{1-D/2}
\label{eq:J3jet}
\end{eqnarray}
with an analogous expression holding also for
$J_{4}^{(b)}(0)$.

For the finite parts of the ``jet'' contribution, one obtains
\begin{equation}
  J_{1}^{(b),\text{fin}}(0) + J_{2}^{(b),\text{fin}}(0)
  + 4 \left[ J_{3}^{(b),\text{fin}}(0) + J_{4}^{(b),\text{fin}}(0) \right]
=
  {1\over 4\pi^{2}} \ln ^{2}
  \left( {\mu^{\prime 2}\over\lambda^{2}} \right)
  - {1\over 8\pi^{2}} \ln
  \left( {\mu^{\prime 2}\over\lambda^{2}} \right).
\label{eq:J3+4fin}
\end{equation}

Collecting our findings from the real-gluon analysis to the
second-order level, we write for the finite contribution to the
cross-section
\begin{equation}
  U_{C\bar{C},S}^{(2)}
=
  1
+
  \frac{\alpha_{\text{s}}}{\pi}C_{\rm F}
  \left[
         \left( \gamma \coth\gamma - 1 \right) \ln
         \left( {\mu^{\prime 2}\over\lambda^{2}}
         \right)
        -{1\over 2}\ln\left( {\mu^{\prime 2}\over\lambda^{2}} \right)
        +\ln ^{2}\left( {\mu^{\prime 2}\over\lambda^{2}} \right)
  \right],
\label{eq:UcuspUcoll}
\end{equation}

At the same time, the singularity structure of the full expression
for the cross-section entails a multiplicative renormalization factor,
which is common to all ``Wilson loop'' configurations entering its
description, but which is the {\it only} one that pertains to the
family of eikonal-type paths under consideration.
The reasoning is, of course, identical to the one given for the
virtual gluon case.
Therefore, the corresponding contribution to the cross-section
factorizes and the same resummation procedure can be employed as
for the virtual-gluon case.
As already observed, the anomalous dimension is in both cases the
same.
There are, however, the following notable differences.
First, the upper limit for the momentum of real-gluon emission is
$1/b^{2}$ instead of $Q^{2}$.
Second, there is a difference of sign, which becomes evident
by comparing Eq.~(\ref{eq:UfinCz}) with Eq.~(\ref{eq:UcuspUcoll}).
Finally, no hard real-gluon emission enters our considerations -
by definition.
In this light, it is practically self-evident that the resummed
expression for real-gluon emission becomes
\begin{equation}
  U_{C\bar{C}}
=
  \exp\left\{ \int_{\bar{\lambda}^{2}}^{c/b^{2}}
  \frac{dt}{t}\left[ \ln{Q^{2}\over t}
               \Gamma_{\text{cusp}}\left( \alpha_{\text{s}}(t) \right)
               -\Gamma\left( \alpha_{\text{s}}(t) \right)
              \right]
      \right\}
              U_{C\bar{C},0},
\label{eq:res_real_gl_emi}
\end{equation}
where
$c=4{\rm e}^{-2\gamma_{\rm E}}$
corresponds to the canonical choice \cite{CSS85}.

We can now bring together real and virtual gluon results by referring
to our generic expression for the cross-section as given by
Eq.~(\ref{eq:wilson}).
First, we write
\begin{equation}
  {\cal W}
=
  \left\langle Tr \left( U^{\dagger} ( \bar{C}^{z^{\prime}})
  U(C^z) \right)
  \right\rangle
=
    U_{C,{\rm cusp}}
    U_{\bar{C},{\rm cusp}}
    U_{C\bar{C},{\rm cusp}}\hat{W}
  + {\cal O}\left( \frac{1}{Q^{2}} \right)
\label{eq:3Us}
\end{equation}
where the factor $\hat{W}$ contains both hard and residual soft
contributions.

Then, at the cross-section level, our threshold resummation
of the virtual gluons reads
\begin{equation}
  U_{\rm C}U_{\bar{\rm C}}
=
  \exp\left\{-\int_{\bar{\lambda}^{2}}^{Q^{2}}
  \frac{dt}{t}
  \left[ \ln{Q^{2}\over t}\Gamma_{\text{cusp}}(\alpha_{\text{s}}(t))
        -\Gamma(\alpha_{\text{s}}(t))
  \right]
      \right\}
              U_{{\rm C},0}U_{\bar{\rm C},0}.
\label{eq:rescrossecvirt}
\end{equation}

Thus, combining the above expressions, the final result reads
\begin{equation}
  {\cal W}
=
  \exp\left\{ -\int_{c/b^{2}}^{Q^{2}}\frac{dt}{t}
  \left[
         \ln{Q^{2}\over t} \Gamma_{{\rm cusp}}(\alpha_{\text{s}}(t))
        -\Gamma(\alpha_{\text{s}}(t))
  \right]
      \right\}{\cal W}_{0},
\label{eq:finres}
\end{equation}
with $\Gamma_{{\rm cusp}}$ and $\Gamma$ given by
Eqs.~(\ref{eq:Gammacusp}) and (\ref{eq:GammaUcoll}),
respectively, an expression obtained before in \cite{KS95},
employing Wilson lines as a quantity attached to quark current
operators.

\section{Concluding remarks}
\label{sec:conclusons}
In this paper we have applied first-quantization techniques to study
threshold resummation of soft gluon radiation for DY-type of processes
in QCD.
We have addressed our efforts in an energy regime whose lower cutoff is
high enough to justify an analysis in which reference to ``gluons'', as
dynamical degrees of freedom, continues to make sense.\footnote{We have
implicitly assumed the pre-confinement property, originally articulated
in the first work of reference \cite{AV79} (see also the second one),
according to which the non-perturbative dynamics responsible for
confinement screens color up to the infrared scale $\lambda$ which sets
the lower limit for the perturbative regime.}

First-quantization approaches to the study of relativistic quantum
systems, involving either strings or world-line agents as their basic
tools, have been employed by a number of authors during the last decade
(the interested reader is recommended to the recent reviews
\cite{ST02,Schu01}) as a viable alternative to the traditional
second-quantization procedure, associated with their field theoretical
casting.
The main feature of this type of approach at the perturbative level, is
that it allows a space-time description of the field theoretical system
leading to expressions that accommodate a host of Feynman diagrams at
once.
On the other hand, it can play a crucial role in the study of
non-perturbative effects \cite{ST02}.

The chief issues of the present methodology can be summarized as
follows:
\begin{itemize}
\item The formalism is based on the Polyakov world-line path integral,
whose particular merit is that it is structured in terms of the spin
factor, a quantity which is path-determined and which accounts
geometrically for the spin of the propagating particle-entity.
Within this framework, the quark-gluon dynamics are embodied in the
expectation value of open/closed Wilson lines.
Thus, quantities of interest for us, such as amplitudes or cross
sections can be solely described in terms of appropriately weighted
integrals over Wilson contours.
\item We have based our considerations on the large scale $Q^2$ in
order to apply the mathematically well-founded fact that the
{\it local} characteristics (like endpoints, cusps, etc.) of the
contours involved in the path integral can be dissected out
(cf. Eq.~(\ref{eq:3Us})).
In this way we calculated the resummed expression for soft-gluon
emission that gives rise to the Sudakov factor.
\item We reiterate that our primary goal has not been to reproduce
known results, but to show {\it how} to obtain them {\it on the
cross-section level} using the world-line Polyakov path integral.
It is obvious that this type of approach can be used to describe
DIS-type processes as well (see \cite{KKS02} from which the present
investigation partly derives).
Furthermore, since we are not obliged to use special type of paths
and we can always stay within the Euclidean formulation -- at least
as long as we care about the leading behavior of quantities like the
cross section, the present investigation may pave the way to extend
this type of approach to the large transverse distance regime, where
we shall meet power corrections signaling nonperturbative
contributions \cite{KS95,CSS85,BB95,SSK00,KS01,Taf01}.
Thus, we hope that this formalism can be extended to the calculation
of the non-perturbative effects of the expectation value of Wilson
loops by having recourse to the extensive existing literature
\cite{ST02,DiDoShSi00} on the subject.
\end{itemize}

\newpage

\appendix
\section{}
Our task is to establish Eqs.~(\ref{eq:IDY}) in
the text.
Performing the integration entering the right hand side of
Eq.~(\ref{eq:virtglex2}), one obtains
\begin{eqnarray}
  I_{3}^{(a)}
&=&
  \frac{1}{4\pi^{2}}\left( -\pi\mu^{2} \right)^{(4-D)/2}
  \Gamma\left( {D\over 2}-1 \right)
  {1\over 4-D}\, {1\over D-3}\, 2w
\nonumber \\
&& \times
  \left\{ w F\left( 1,{D\over 2}-1;{D-1\over 2};1-w^{2} \right)
  \right.
\nonumber \\
&&
\left.
  +{1\over 2}\left[ 2(1-w) \right]^{2-D/2}
   F\left( 1,{D\over 2}-1;{D-1\over 2};{1+w\over 2} \right)
 \right\}.
\label{eq:A.1}
\end{eqnarray}

Setting $D=4$, we obtain
\begin{equation}
  F\left( 1,1;3/2;1-w^{2} \right)
=
  \frac{\gamma}{\sinh\gamma\cosh\gamma}
\end{equation}
and
\begin{equation}
   F\left( 1,1;3/2;{1+w\over 2} \right)
=
   \frac{\gamma}{\sinh\gamma}
  -i\frac{\pi}{\sinh\gamma}.
\end{equation}
As the imaginary part in the above expression will cancel against its
counterpart in the conjugate expression, it can be dropped as far as the
cross-section is concerned.

Denoting the expression inside the curly brackets on the rhs
of Eq.~(\ref{eq:A.1}) by $f_{D}(w)$ and setting
\begin{equation}
    f_{D}(w)
=
    f_{4}(w)+(4-D)\frac{f_{D}(w) - f_{4}(w)}{4-D},
\end{equation}
we realize that the second term on the rhs will lead to finite terms
that depend solely on $w$ and which will cancel against similar
contributions of the same sort coming from the other terms entering
Eq.~(\ref{eq:U2order}).
Putting everything together, one finally arrives at
Eq.~(\ref{eq:IDY}).

To establish the result given by Eq.~(\ref{eq:I3fin}), we first
note that Eq.~(\ref{eq:I3}) gives
\begin{eqnarray}
  I_{3}^{(b)}
&=&
  \frac{1}{4\pi^{2}}
  \left( -\pi{\mu^{2}\over\lambda^{2}} \right)^{(4-D)/2}
  \Gamma\left( {D\over 2}-1 \right)
  {1\over (4-D)^{2}}
  \left( \frac{2v\cdot v^{\prime}}{|v|^{2}} \right)^{(4-D)/2}
\nonumber\\
&& \times
   \left[
         F\left( {D\over 2}-1,2-{D\over 2};
                3-{D\over 2};-{2v\cdot v^{\prime}\over |v|^2}
          \right)
       + \left( 1+{2v\cdot v^{\prime}\over |v|^{2}}
       \right)^{2-D/2}\right.
\nonumber \\
&&
\left.
      \quad\;\; - \left( {2v\cdot v^{\prime}\over |v|^{2}} \right)^{2-D/2}
   \right].
\end{eqnarray}

Then, in the limit $D\rightarrow 4$ one easily retrieves Eq.~(\ref{eq:I3fin}).

\end{document}